\documentclass[pra,twocolumn,floats,showpacs,preprintnumbers,tighten,epsfig,superscriptdaddress,raggedbottom]{revtex4}

\usepackage{graphicx}
\usepackage{dcolumn}
\usepackage{bm}
\usepackage{amsmath}
\usepackage{amsfonts}
\usepackage{amssymb}

\usepackage{bm,color}

\begin{document}

\title{Optomechanical tests of a Schr\"{o}dinger-Newton equation for gravitational quantum mechanics}

\author{C.~ C.~Gan}
\affiliation{
Department of Quantum Science, Research School of Physics and Engineering, Australian National University, Canberra ACT 0200, Australia}
\affiliation{Department of Physics, University of Malaya, 50603 Kuala Lumpur, Malaysia}
\author{C.~M.~Savage} 
\email{craig.savage@anu.edu.au}
\author{S.~Z.~Scully} 
\affiliation{
Department of Quantum Science, Research School of Physics and Engineering, Australian National University, Canberra ACT 0200, Australia}

\date{\today {}}

\begin{abstract}
We show that optomechanical systems can test the Schr\"{o}dinger-Newton equation of gravitational quantum mechanics due to Yang \textit{et al.} \cite{Yang}. This equation is motivated by semiclassical gravity, a widely used theory of interacting gravitational and quantum fields. From the many-body Schr\"{o}dinger-Newton equation follows an approximate equation for the center-of-mass dynamics of macroscopic objects. This predicts a distinctive double-peaked signature in the output optical quadrature power spectral density of certain optomechanical systems. Since the Schr\"{o}dinger-Newton equation lacks free parameters, these will allow its experimental confirmation or refutation.
\end{abstract}

\maketitle

\section{Introduction}
\label{Introduction}

There are many candidate theories for gravitational quantum mechanics - that is for the interaction between gravity and quantum mechanical systems \cite{Carlip, Kafri, Anastopoulos2}. Theories in which the gravitational field is quantized are notoriously difficult to test experimentally because their consequences emerge at the relatively inaccessible Planck scale \cite{Amelino-Camelia, Lammerzahl}. Nevertheless, some theories of quantum gravity have been tested by cosmological observations. For example, spacetime-foam theories predict a variation of the speed of light with frequency that is constrained by x-ray and gamma-ray sources \cite{Vasileiou, Perlman}; but see Chen \textit{et al}. \cite{Chenfoam}.  Other theories may be experimentally testable with laboratory experiments \cite{Yang, Yang2}. In particular, modifications to position-momentum commutation relations \cite{Ali} may be testable using optomechanics \cite{Amelino, Pikovski, Hogan, Bawaj, Berchera, Belenchia, Albrecht, Bowen} or gravitational wave bar detectors \cite{Marin2014}. 

This paper models possible experimental tests of the specific theory of gravitational quantum mechanics expressed by the many-body Schr\"{o}dinger-Newton equation \cite{Penrose,Diosi,Yang,Giulini, Colin, Bahrami}. This equation is motivated by semiclassical gravity \cite{Rosenfeld, Kiefer book, Kiefer ISRN}, for which the quantum mechanical expectation value of the stress-energy-momentum tensor operator $\hat{T}_{\mu \nu}$ is the source of the Einstein tensor $G_{\mu \nu}$,
\begin{equation} \label{SCGE}
G_{\mu \nu} = \tfrac{8 \pi G}{c^4} \langle \hat{T}_{\mu \nu}  \rangle ,
\end{equation}
where $G$ is the Newtonian gravitational constant, and $c$ is the speed of light. Semiclassical gravity belongs to the class of unification theories in which gravity is a classical field \cite{Albers, Hu, Bahrami2, Carlip2}. Another type of possible experimental test of classical gravity would investigate whether or not the gravitational interaction can transmit quantum coherence \cite{Kafri}.   The relationship of the many-body Schr\"{o}dinger-Newton equation to semiclassical gravity is discussed in detail by Hu \cite{Hu2014}.

Due to the difficulties of coupling classical and quantum theories there are reasons to doubt that Eq.~(\ref{SCGE}) is true at a fundamental level \cite{Kapustin}. Furthermore, in the absence of a complete quantum theory of gravity, its domain of validity is unclear \cite{Hu}. Nevertheless, it has long been a useful approximation for studying quantum effects on gravitational fields, such as the back-reaction from Hawking radiation.  

Page and Geilker \cite{Page experiment} concluded from a Cavendish type experiment that semiclassical gravity does not describe macroscopic quantum superpositions. This conclusion depends on the validity of their manual procedure for putting kilogram masses into a quantum mechanical superposition state, which is contentious given our understanding of decoherence \cite{Mattingly, Blencowe}. The analysis in this paper is restricted to Gaussian states and hence does not apply to position superposition states.

In the next section the Schr\"{o}dinger-Newton equation for an oscillator's center-of-mass is discussed. In section \ref{Optomechanical theory} it is included in a simple optomechanical model and expressions for the quadrature power spectral densities of the output light are derived. In section \ref{The semiclassical signature in optomechanical systems} these are evaluated for parameters corresponding to three specific systems: a kilogram scale torsion pendulum, a somewhat more massive Laser Interferometer Gravitational-wave Observatory (LIGO) like system, and a milligram scale levitating mirror. We conclude with an evaluation of the experimental prospects for testing the Schr\"{o}dinger-Newton equation.

\section{The Schr\"{o}dinger-Newton equation}
\label{SN equation}

The Newtonian limit of semiclassical gravity motivated Penrose \cite{Penrose} and Di\'{o}si \cite{Diosi} to postulate that the gravitational potential energy in the many-body Schr\"{o}dinger equation should depend on the quantum mechanical mass density, and hence on the wavefunction. The resulting equation is known as the many-body Schr\"{o}dinger-Newton equation. 

Yang \textit{et al.}~\cite{Yang}, and others \cite{Giulini, Colin, Bahrami}, have derived from the many-body Schr\"{o}dinger-Newton equation an approximate equation for the wavefunction $\Psi (x,t)$ of the center-of-mass degree of freedom of a macroscopic mechanical oscillator,
\begin{align} \label{SNE}
i \hbar \frac{\partial \Psi (x,t)} {\partial t} =& \left[ -\tfrac{ \hbar^2 }{2M} \partial_x^2 +\tfrac{M}{2} \omega_0^2 x^2
  \right. \nonumber \\
& \left . + \tfrac{M}{2}  \omega_{\textrm{SN}}^2 (x- \langle x  \rangle)^2 - F x \right] \Psi (x,t).
\end{align}
$\omega_0$ is the mechanical oscillator's angular frequency and $M$ its mass. $x$ is the small deviation of the center-of-mass from its steady-state equilibrium position. $F$ represents zero-mean classical driving forces, large enough that their quantum mechanical aspects are negligible; for example, thermal noise from the mirror suspension. In this paper we analyze potential optomechanical tests of this equation.

The Schr\"{o}dinger-Newton equation (\ref{SNE}) is nonlinear in the wavefunction because of the position  expectation value $\langle x  \rangle$ that arises from the expectation value in Eq.~(\ref{SCGE}). It is due to gravitational self-interaction \cite{Yang, Giulini, Colin, Bahrami} and is proportional to the square of the Schr\"{o}dinger-Newton frequency, $ \omega_{\textrm{SN}}^2$, discussed below. Yang \textit{et al.}~\cite{Yang} realized that the relevant density scale for gravitational self-interaction is not  the density of the material, but rather the much larger density of nuclei. Therefore, the dynamical effect on optomechanical systems might be large enough to be measureable. It has also been suggested that the effect on the energy spectrum of an optically trapped microdisk might be observable \cite{Grossardt}.

Nonlinear Schr\"{o}dinger equations may arise from quantum field theory as descriptions of the dynamics of condensate wavefunctions. However, the Schr\"{o}dinger-Newton equation is for an uncondensed system. If Eq.~(\ref{SNE}) were found to be an accurate description of experimental systems this would imply that the gravitational interaction was fundamentally different to  all the other interactions. In particular, self-interaction would not be eliminated by renormalization, as it is for the electromagnetic field \cite{Anastopoulos}. Some of the difficulties created by nonlinear Schr\"{o}dinger equations, such as faster than light signalling, may be overcome by the addition of stochastic terms \cite{Nimmrichter}. However, nonlinear quantum mechanics may require new theories of measurement and interpretations of the wavefunction \cite{Weinberg}. In this paper we assume that the theory of Yang \textit{et al.}~\cite{Yang} is a useful approximation to the more complete theories being developed \cite{YanbeiMeasureableSignatures}.

Assuming Gaussian wavefunctions for the nuclei positions, the squared Schr\"{o}dinger-Newton frequency is approximately \cite{SNfrequencyderivation,Iwe},
\begin{equation} \label{SNfrequency}
 \omega_{\textrm{SN}}^2 \approx  \frac{G m}{12 \sqrt{\pi} \Delta x_\textrm{nuc}^3} .
\end{equation}
In the Debye approximation, the position variance of a nucleus at temperature $T$ is \cite{Willis},
\begin{equation} \label{Debyeapprox}
\Delta x_\textrm{nuc}^2 = \frac{3 \hbar^2}{4 m k_b \Theta_D^2} \left[ \Theta_D + 4T \Phi (\Theta_D / T)  \right] ,
\end{equation}
where $k_b$ is Boltzman's constant, $\hbar$ is the reduced Planck's constant, $m$ is the nuclear mass, $\Theta_D$ is the Debye temperature of the material, and $\Phi (x)$ is the Debye integral function,
\begin{equation} \label{Debyeintegralfunction}
\Phi (x) = \frac{1}{x} \int_0^x \frac{y}{e^y-1} dy .
\end{equation}
Near zero temperature, $T \ll \Theta_D$, the spread of the nuclei is then the zero-point value,
\begin{equation} \label{DebyeApproxZeroT}
\Delta x_{\textrm{nuc,}T=0}^2 = \frac{3 \hbar^2}{4 m k_b \Theta_D},
\end{equation}
and the squared Schr\"{o}dinger-Newton frequency is,
\begin{equation} \label{SNfrequencyZeroT}
\omega_{\textrm{SN}}^2 \approx
\frac{2 G}{9 \sqrt{3 \pi} \hbar^3} m^\frac{5}{2} ( k_b \Theta_D )^\frac{3}{2} .
\end{equation}
This increases with the nuclear mass and with the confinement of the nuclei by the solid, as represented by the Debye temperature. 

Eq.~(\ref{SNE}) arises from an expansion of the gravitational interaction energy in the ratio $\Delta x_\textrm{cm} / \Delta x_\textrm{nuc}$, where $\Delta x_\textrm{cm}$ is the uncertainty in the oscillator's center-of-mass position \cite{Yang}. Hence its validity requires $\Delta x_\textrm{cm} / \Delta x_\textrm{nuc} \ll 1$. If we assume that the center-of-mass is in thermal equilibrium, its effective  temperature, $T_\textrm{cm}$, follows from equipartition of energy, so that the average number of center-of-mass vibrational quanta $n$ is estimated from $k_b T_\textrm{cm} = n \hbar \omega_0$. The corresponding center-of-mass variance is $\Delta x_\textrm{cm}^2 \approx  (2n+1) \hbar /(2M \omega_0) \approx k_b T_\textrm{cm} / (M \omega_0^2)$. Using this and Eq.~(\ref{DebyeApproxZeroT}) in $( \Delta x_\textrm{cm} / \Delta x_{\textrm{nuc,}T=0} )^2 \ll 1$ gives,
\begin{equation} \label{TemperatureInequality}
T_\textrm{cm} \ll \frac{3M \hbar^2 \omega_0^2}{4 m k_b^2 \Theta_D}  .
\end{equation}
For silicon this becomes $T_\textrm{cm} \ll (1.3 \textrm{ K kg}^{-1}\textrm{s}^{-2}) M \omega_0^2$ and for iron, $T_\textrm{cm} \ll (0.9 \textrm{ K kg}^{-1}\textrm{s}^{-2}) M \omega_0^2$. Hence for low-mass, or low-frequency, systems the center-of-mass mode may require cooling \cite{Miao} to ensure the validity of the Schr\"{o}dinger-Newton equation (\ref{SNE}).

For a compound such as quartz (SiO$_2$), we estimate the Schr\"{o}dinger-Newton frequency to be the mass fraction weighted quadrature sum of the Schr\"{o}dinger-Newton frequencies of the elemental components. This is because the total gravitational self-interaction energy is the sum of that due to each nucleus, and hence each element contributes to it independently \cite{Yang}. For example for quartz,
\begin{equation} \label{SNquartz}
\omega_\textrm{SN}^2 = f_\textrm{O}  \, \omega_\textrm{SN,O}^2 + f_\textrm{Si} \,  \omega_\textrm{SN,Si}^2  \, ,
\end{equation}
where $f_\textrm{O} = 0.53$ and $f_\textrm{Si} = 0.47$ are the fractions of the oscillator mass due to oxygen and silicon respectively. We estimate the Schr\"{o}dinger-Newton frequencies for each element, $\omega_\textrm{SN,O}$ and $\omega_\textrm{SN,Si}$, using the Debye temperature for quartz, $\Theta_D = 470$ K, giving: $\omega_\textrm{SN} \approx 0.027$ s$^{-1}$ at $T = 0$ K and $\omega_\textrm{SN} \approx 0.013$ s$^{-1}$ at $T = 293$ K. 

Table \ref{Materials} lists the Schr\"{o}dinger-Newton frequencies of various elements. Osmium has the highest due to its high nuclear mass. Note that for fluids, such as room temperature mercury, the nuclei are weakly confined and their Schr\"{o}dinger-Newton frequency is effectively zero. The material dependence of the Schr\"{o}dinger-Newton frequency could be useful for confirmation of any experimental results supporting the Schr\"{o}dinger-Newton equation.

\begin{table}
\caption{\label{Materials}Schr\"{o}dinger-Newton frequencies, atomic masses, and Debye temperatures of selected elements. The frequencies, with units of s$^{-1}$, are evaluated at 0 K and 293 K. The Carbon entry is for diamond.}
\begin{ruledtabular}
\begin{tabular}{lrrrr}
Element & $\omega_\textrm{SN}$  (0 K) & $\omega_\textrm{SN}$  (293 K) & $m$ (amu) & $\Theta_D$ (K) \\
\hline
Be & 0.016 & 0.012 & 9 & 1140\\
Al & 0.032 & 0.027 & 28 & 428\\  
C & 0.038 & 0.036 & 12 & 2230\\ 
Si & 0.043 & 0.025 & 28 & 645\\
Fe & 0.082 & 0.039 & 56 & 470\\
Nb & 0.10 & 0.034 & 92 & 275\\
Os & 0.38 & 0.18 & 190 & 500\\
\end{tabular}
\end{ruledtabular}
\end{table}

In the next section we show that when the oscillator is optomechanically coupled to light, the optical quadrature power spectral density has two resonant features \cite{Yang}. For a resonant cavity with zero detuning, one is at the oscillator frequency $\omega_0$, and the other at the higher semiclassical frequency,
\begin{equation} \label{semiclassical frequency approximation}
 \omega_{\textrm{sc}} = ( \omega_0^2 + \omega_{\textrm{SN}}^2 )^\frac{1}{2}. 
\end{equation}
For non-zero cavity detuning these frequencies are shifted by the optomechanical spring effect. This double resonance is the signature of the Schr\"{o}dinger-Newton equation.

When the oscillator frequency is much greater than the Schr\"{o}dinger-Newton frequency, $\omega_0 \gg \omega_{\textrm{SN}}$, the semiclassical frequency is approximately,
\begin{equation} \label{semiclassical frequency approximation}
\omega_{\textrm{sc}} \approx
\omega_0 + \tfrac{1}{2} \omega_{\textrm{SN}} \left( \frac{\omega_{\textrm{SN}}}{\omega_0} \right) .
\end{equation}
In this limit, separation of the features requires a mechanical $Q_\textrm{m}$ of at least $( \omega_0 / \omega_{\textrm{SN}})^2$. For example, an oscillator made of quartz with $\omega_0 \approx 10 \, \textrm{s}^{-1}$ requires $Q_\textrm{m} > 10^6$, which is expected to be achievable for a levitating mirror \cite{Guccione}. However, for higher frequency oscillators the required $Q_\textrm{m}$ is beyond what is currently achievable, e.g. $Q_\textrm{m} > 10^{16}$ for $\omega_0 \approx 10^6 \, \textrm{s}^{-1}$. Hence, due to the sub-Hertz values of the Schr\"{o}dinger-Newton frequency and the inverse  dependence of $\omega_\textrm{sc} - \omega_0$ on the oscillator frequency, low-frequency systems are the most promising for testing semiclassical gravity.

\section{Optomechanical theory}
\label{Optomechanical theory}

Since the potential in the Schr\"{o}dinger-Newton equation (\ref{SNE}) is quadratic, it preserves the Gaussian wavefunction form under time evolution \cite{Tannor}. Hence for Gaussian wavefunctions equivalent equations may be found for the means and covariances of the centre-of-mass position $\hat{X}$ and momentum $\hat{P}$ operators. Following Yang \textit{et al.} \cite{Yang}, these also follow from effective Heisenberg equations for $\hat{X}$ and $\hat{P}$,
\begin{equation} \label{Heisenberg oscillator}
\frac{d \hat{X}}{dt} = \frac{\hat{P}}{M}, \,
\frac{d \hat{P}}{dt} = -M \omega_0^2 \hat{X} -M \omega_{\textrm{SN}}^2 (\hat{X} -\langle \hat{X} \rangle ) +F .
\end{equation}
Although these are only valid for Gaussian wavefunctions, they are convenient for optomechanics because they facilitate coupling to the quantized optical field.

\begin{figure}[b]
\includegraphics[width=6cm]{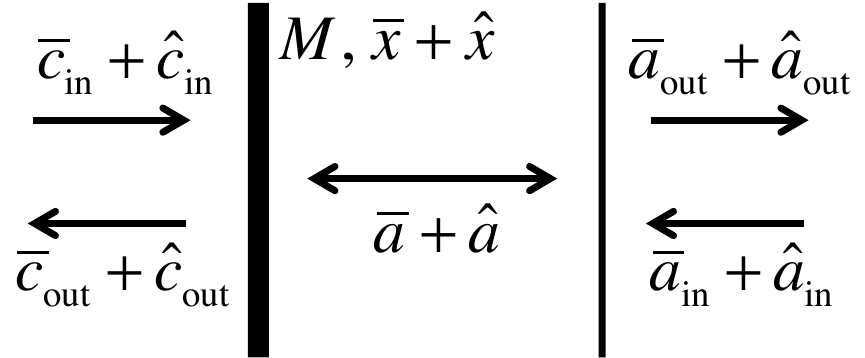}
\caption{Schematic diagram of the optomechanical system. The heavy line represents the harmonically confined mirror of mass $M$ with position $\bar{x}+\hat{x}$, that is positive to the left. The other line is the fixed partially transmitting input/output mirror. The field in the cavity is $\bar{a}+\hat{a}$. The cavity input and output fields are $\bar{a}_{\textrm{in}}+\hat{a}_{\textrm{in}}$ and $\bar{a}_{\textrm{out}}+\hat{a}_{\textrm{out}}$. The probe input and output fields are $\bar{c}_{\textrm{in}}+\hat{c}_{\textrm{in}}$ and $\bar{c}_{\textrm{out}}+\hat{c}_{\textrm{out}}$. }
\label{cavity}
\end{figure}

We analyze the case of the mechanical oscillator forming one totally reflecting mirror of an optical cavity, Fig.~\ref{cavity}. We assume a single cavity mode with classical amplitude $\bar{a}$, and a perturbative quantum field with annihilation operator $\hat{a}$, so that the total cavity field is $\bar{a} + \hat{a}$, which is dimensionless. The input field $\bar{a}_{\textrm{in}} +\hat{a}_{\textrm{in}}$ enters through the second fixed mirror through which the output field $\bar{a}_{\textrm{out}} +\hat{a}_{\textrm{out}}$ also exits.

We also include an additional probe field reflecting directly off the other side of the free mirror \cite{Schmole}; this was the case considered by Yang \textit{et al.} \cite{Yang}. These incident and reflected probe fields are denoted: $\bar{c}_{\textrm{in}} +\hat{c}_{\textrm{in}}$ and $\bar{c}_{\textrm{out}} +\hat{c}_{\textrm{out}}$.
These have the dimensions s$^{-\frac{1}{2} }$ and are also split into the sum of a classical part, denoted by an overbar, and a perturbative quantum field. 

The input fields have delta-function commutation relations, $[\hat{a}_\textrm{in} (t) , \hat{a}_\textrm{in}^\dagger (t')] = [\hat{c}_\textrm{in} (t) , \hat{c}_\textrm{in}^\dagger (t')] = \delta(t-t')$. They may be regarded as unitarily transformed by coherent state displacement operators \cite{Bowen, WallsandMilburn} so that their classical parts are coherent amplitudes, their expectation values are zero, and if the inputs are coherent states, the transformed states are vacuums.

The mirror position operator is also separated into its classical part $\bar{x}$ and a perturbative quantum part $\hat{x}$, so that $\hat{X} = \bar{x} + \hat{x}$. The classical parts of the cavity field $\bar{a}$ and mirror position $\bar{x}$ are found by solving the classical optomechanical equations for the steady state:
\begin{align} \label{Heisenberg means}
& M  \omega_0^2 \bar{x} = \hbar \omega_c \bar{a}^2 /L  -2 \hbar k \bar{c}_\textrm{in}^2 , \nonumber  \\
& (\gamma_c /2 -i (\omega_l - \omega_c)) \bar{a} = i \omega_c \bar{a} \bar{x}  /L + \sqrt{\gamma_c}  \, \bar{a}_{\textrm{in}} ,
\end{align}
where $\omega_c$ is the cavity resonance frequency, $\omega_l$ is the driving laser frequency,  $k = \omega_l /c$, $\gamma_c$ is the cavity damping, and $L$ the cavity length.
These follow from the Heisenberg equations by taking expectation values, and factorizing expectation values of operator products into products of operator expectation values.

Following Aspelmeyer \textit{et al.}~\cite{Aspelmeyer RMP} the Heisenberg equations to first order in the perturbative quantum operators are:
\begin{align} \label{Heisenberg full}
M ( \frac{d^2 \hat{x}}{dt^2} + \gamma_m \frac{d \hat{x}}{dt} + \omega_0^2 \hat{x} 
+ \omega_{\textrm{SN}}^2 (\hat{x} -\langle \hat{x} \rangle ) ) &=   \nonumber  \\
\hbar G_\textrm{a} (\hat{a} +\hat{a}^\dagger) - \hbar G_\textrm{p} (\hat{c}_\textrm{in} +\hat{c}_\textrm{in}^\dagger) &+ F, \nonumber  \\
\frac{d \hat{a}}{dt} +(\gamma_c /2 -i \Delta) \hat{a} = i G_\textrm{a} \hat{x} + \sqrt{\gamma_c}  \, \hat{a}_{\textrm{in}} , \nonumber  \\
\hat{c}_{\textrm{out}} = - \hat{c}_{\textrm{in}} -  i G_\textrm{p} \hat{x} ,
\nonumber  \\
 \hat{a}_{\textrm{out}} = \hat{a}_{\textrm{in}} -\sqrt{\gamma_c} \, \hat{a}.
\end{align}
$G_\textrm{a} = | \bar{a} | \omega_c /L$ is the cavity optomechanical coupling strength,  
$G_\textrm{p} = | \bar{c}_\textrm{in} | 2 k$ is the normal incidence probe optomechanical coupling strength. The classical probe input field is given in terms of the input power $P_\textrm{c}$ by, $| \bar{c}_\textrm{in} | = \sqrt{P_\textrm{c} / (\hbar \omega_l)}$, and similarly for the cavity input.
$\gamma_m$ is the mechanical damping. 
The cavity detuning $\Delta = \omega_l - \omega_c - (\omega_c / L) \bar{x}$, includes that due to the shift of the mirror position to $\bar{x}$ by the classical light pressure \cite{Bowen, Aspelmeyer RMP}. 

The Fourier transforms of the Heisenberg equations (\ref{Heisenberg full}) are linear algebraic equations for which we present analytic solutions. We denote Fourier transforms with a tilde: $\tilde{z} (\omega) = \int_{-\infty}^{\infty} e^{i \omega t} z(t) dt$. In our notation $\hat{\tilde{a}}^\dagger (\omega)$ is the Fourier transform of $\hat{a}^\dagger (t)$ and therefore 
$( \hat{\tilde{a}} (\omega) )^\dagger = \hat{\tilde{a}}^\dagger (-\omega)$. In the Fourier domain the commutation relations are then: $[\hat{\tilde{a}}_\textrm{in} (\omega) , \hat{\tilde{a}}_\textrm{in}^\dagger (\omega')] = [\hat{\tilde{c}}_\textrm{in} (\omega) , \hat{\tilde{c}}_\textrm{in}^\dagger (\omega')] =2 \pi \delta(\omega+\omega')$.

Following the the supplemental material for \cite{Yang}, taking the expectation values of the Fourier transformed Heisenberg equations we can solve for the expectation value of the position Fourier transform,
\begin{equation} \label{x mean}
\langle \hat{\tilde{x}} \rangle = \chi_\textrm{m} \tilde{F}  ,
\end{equation}
where the susceptibility $\chi_\textrm{m}$ is defined below in Eqs.~(\ref{transfer functions}). Hence the mean perturbative oscillator position only responds to the classical force. There is no effect from the gravitational self-interaction.

The expression for $\langle \hat{\tilde{x}} \rangle$ may be substituted into the Fourier transforms of Eqs.~(\ref{Heisenberg full}) and solved for the Fourier components of the cavity output field in terms of the input fields,
\begin{equation} \label{output form}
\hat{\tilde{a}}_{\textrm{out}}   =  k_{a,1} \hat{\tilde{a}}_{\textrm{in}}  + k_{a,2} \hat{\tilde{a}}_{\textrm{in}}^\dagger 
+ k_{a,3} ( \hat{\tilde{c}}_{\textrm{in}} + \hat{\tilde{c}}_{\textrm{in}}^\dagger )
+ k_{a,4} \tilde{F} ,
\end{equation}
where the coefficients are:
\begin{align} \label{output field coefficients}
k_{a,1}  &=   
 1 - i \gamma_\textrm{c}  \chi_\textrm{cav} (\omega) +i \gamma_\textrm{c} \hbar G_\textrm{a}^2 
 \chi_\textrm{cav}^2 (\omega)   \chi_\textrm{sc} (\omega), \quad
\nonumber  \\ 
k_{a,2} &=  
-i \gamma_\textrm{c}  \hbar G_\textrm{a}^2 \chi_\textrm{cav} (\omega)
\chi_\textrm{cav}^* (-\omega) \chi_\textrm{sc} (\omega),  
\nonumber   \\
k_{a,3}  &= 
-\sqrt{\gamma_\textrm{c}} \hbar G_\textrm{a} G_\textrm{p} 
\chi_\textrm{cav} (\omega) \chi_\textrm{sc} (\omega) ,
\nonumber \\
k_{a,4} &=
\sqrt{\gamma_\textrm{c}} G_\textrm{a}  \chi_\textrm{cav} (\omega) \chi_\textrm{m} (\omega).
\end{align}
We have introduced the cavity response function $\chi_\textrm{cav}$, and susceptibility functions associated with the semiclassical frequency $\chi_\textrm{sc}$, and with the oscillator frequency $\chi_\textrm{m}$, and incorporated the optical spring effect through $\Sigma (\omega)$:
\begin{align} \label{transfer functions}
\chi_\textrm{cav} (\omega) &=
[ (\omega +\Delta) + i \gamma_\textrm{c} /2 ]^{-1}, \nonumber \\
\chi_\textrm{sc} (\omega) & = 
[ M (\omega_\textrm{sc}^2 -\omega^2 ) -i M \gamma_\textrm{m} \omega + \Sigma (\omega)]^{-1} , 
\nonumber \\
\chi_\textrm{m} (\omega) & = 
[ M (\omega_0^2 -\omega^2 ) -i M \gamma_\textrm{m} \omega + \Sigma (\omega)]^{-1} , 
\nonumber \\
\Sigma (\omega) &= 
\hbar G_\textrm{a}^2 ( \chi_\textrm{cav} (\omega) + \chi_\textrm{cav}^* (-\omega) ) .
\end{align}
For the probe field we have,
\begin{align} \label{probe output form}
\hat{\tilde{c}}_{\textrm{out}}   = &  k_{c,1} \hat{\tilde{a}}_{\textrm{in}}  + k_{c,2} \hat{\tilde{a}}_{\textrm{in}}^\dagger 
- \hat{\tilde{c}}_{\textrm{in}} +k_{c,3} ( \hat{\tilde{c}}_{\textrm{in}} + \hat{\tilde{c}}_{\textrm{in}}^\dagger )
\nonumber  \\ 
+ & k_{c,4} \tilde{F} ,
\end{align}
where the coefficients are:
\begin{align} \label{probe output field coefficients}
k_{c,1}  &=   
 \sqrt{\gamma_\textrm{c}} \hbar G_\textrm{a} G_\textrm{p} 
 \chi_\textrm{cav} (\omega) \chi_\textrm{sc} (\omega) , \quad
\nonumber  \\ 
k_{c,2} &=  
- \sqrt{\gamma_\textrm{c}} \hbar G_\textrm{a} G_\textrm{p} 
\chi_\textrm{cav}^* (-\omega) \chi_\textrm{sc} (\omega) ,  
\nonumber   \\
k_{c,3}  &= 
- i \hbar G_\textrm{p}^2 \chi_\textrm{sc} (\omega) ,
\nonumber \\
k_{c,4} &=
-i G_\textrm{p} \chi_\textrm{m} (\omega) .
\end{align}

The above reduce to the conventional optomechanical equations when the Schr\"{o}dinger-Newton frequency is zero, so that $\omega_\textrm{sc} = \omega_0$, and $\chi_\textrm{sc} = \chi_\textrm{m}$. Note that the semiclassical frequency appears in the coefficients of the input quantum fields and that the oscillator frequency only appears in the coefficient of the classical force. Hence, the oscillator responds at the semiclassical frequency to quantum optical noise, and at the oscillator frequency to classical noise. Matsumoto \textit{et al.} \cite{Matsumoto} demonstrated experimentally that the quantum driving of an oscillator can dominate its thermal (classical) driving.

Consider the output field quadratures:
\begin{align} \label{quadratures}
\hat{\tilde{X}}_{\textrm{a}, \theta} (\omega) = \frac{1}{\sqrt{2}} \left( e^{-i \theta} \hat{\tilde{a}}_{\textrm{out}} (\omega) 
+ e^{i \theta} \hat{\tilde{a}}_{\textrm{out}}^\dagger ( \omega) \right) ,
\\
\hat{\tilde{X}}_{\textrm{c}, \theta} (\omega) = \frac{1}{\sqrt{2}} \left( e^{-i \theta} \hat{\tilde{c}}_{\textrm{out}} (\omega) 
+ e^{i \theta} \hat{\tilde{c}}_{\textrm{out}}^\dagger (\omega) \right) ,
\end{align}
where $\theta$ is the quadrature angle.
If the perturbative quantum input field is in the vacuum state then the only non-zero field contributions to the quadrature spectral variance are from $\langle \hat{\tilde{a}}_\textrm{in} (\omega)  \hat{\tilde{a}}_{\textrm{in}}^\dagger (\omega') \rangle = \langle \hat{\tilde{c}}_\textrm{in} (\omega)  \hat{\tilde{c}}_{\textrm{in}}^\dagger (\omega') \rangle = 2 \pi \delta(\omega + \omega')$, so that,
\begin{align} \label{spectral density}
 \langle & \hat{\tilde{X}}_{i,\theta} (\omega) \hat{\tilde{X}}_{i,\theta} (\omega') \rangle = 
 S_{i,\theta} (\omega) 2 \pi \delta(\omega + \omega') 
\nonumber  \\
&  =  2 \pi \delta(\omega + \omega') \times  \tfrac{1}{2} \left( 
 \left| e^{-i \theta} k_{i,1}(\omega) +e^{i \theta} k_{i,2}^*(-\omega) \right|^2   \right.
\nonumber  \\
& +  \left| e^{-i \theta} ( k_{i,3}(\omega) - \delta_{i,\textrm{c}}) +e^{i \theta} k_{i,3}^*(-\omega) \right|^2 
\nonumber  \\
 &  \left. + \left| e^{-i \theta} k_{i,4}(\omega) +e^{i \theta} k_{i,4}^*(-\omega) \right|^2 S_\textrm{cl} 
 \right) 
\nonumber  \\
 &  = \pi \delta(\omega + \omega') \times \left( S_{i,\theta}^\textrm{q} (\omega) + S_{i,\theta}^\textrm{cl} (\omega) \right)  ,
\end{align}
where this defines the dimensionless power spectral density $S_{i,\theta} (\omega)$, normalized to shot noise, for either the cavity output field $i=\textrm{a}$ or the probe field $i=\textrm{c}$.  $S_\textrm{cl}$ is the power spectrum of the classical noise. In the last line we have separated the spectral power into the part $S_{i,\theta}^\textrm{cl} (\omega)$, proportional to $S_\textrm{cl}$, and the rest, $S_{i,\theta}^\textrm{q} (\omega)$, which is driven by quantum optical noise. We only consider thermal driving of the mirror suspension \cite{Bowen},
\begin{equation} \label{thermal power spectrum}
S_\textrm{cl} = 2 k_b T_\textrm{s}  M \gamma_\textrm{m} ,
\end{equation}
where $T_\textrm{s}$ is the temperature of the suspension.  For convenience we refer to $S_{i,\theta}^\textrm{cl} (\omega)$ and $S_{i,\theta}^\textrm{q} (\omega)$ as the thermal and quantum contributions respectively.

The signal is strongest in the phase quadratures of the probe and cavity output fields. There is no signal in the amplitude quadrature. This is because for the probe the mirror motion causes path length changes only, and for the lossless cavity we are assuming, the cavity output intensity does not change. Hence we now limit our analysis to the phase quadrature, $\theta = \pi/2$.

\section{The Schr\"{o}dinger-Newton signature in optomechanical systems}
\label{The semiclassical signature in optomechanical systems}
\begin{figure}[b]
\includegraphics[width=\linewidth]{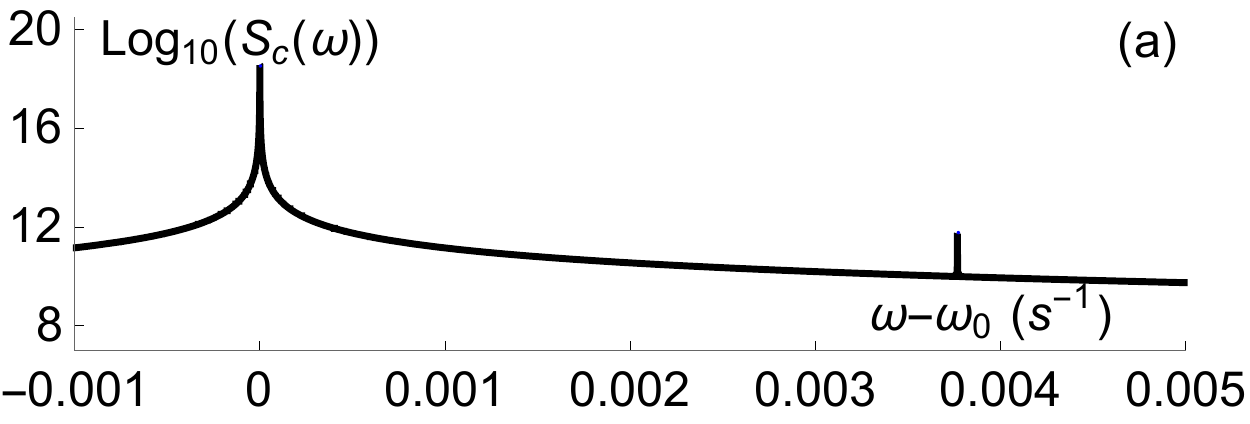}
\\
\vspace{3mm}
\includegraphics[width=\linewidth]{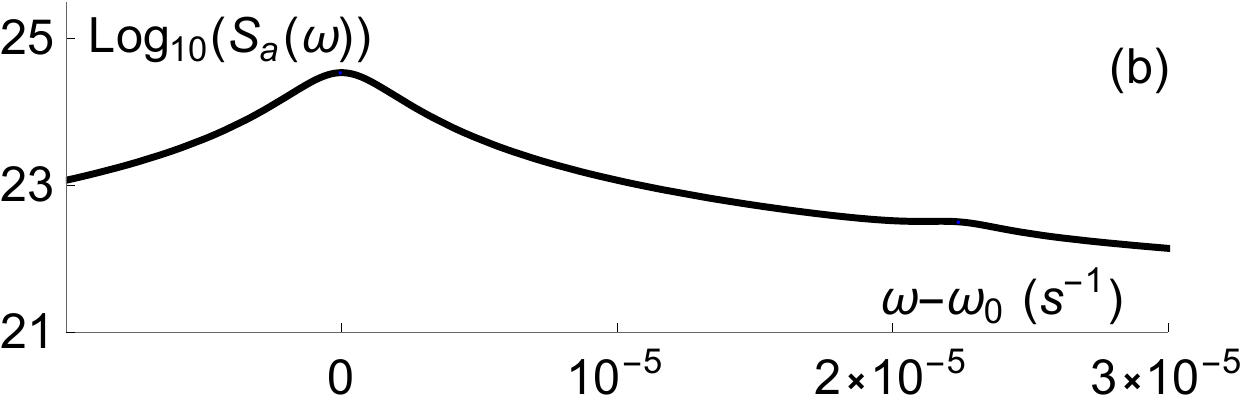}
\\
\includegraphics[width=\linewidth]{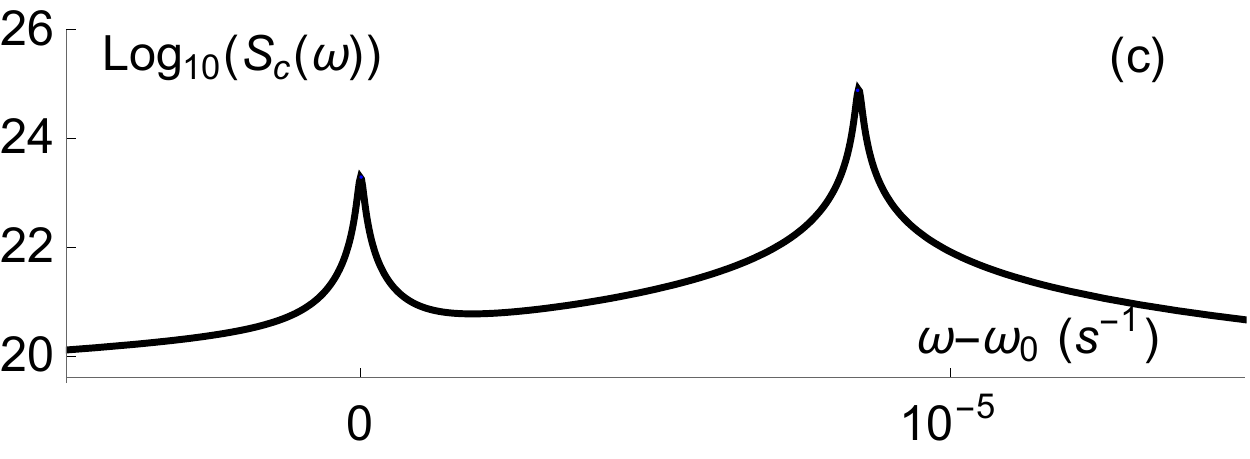}
\caption{Logarithmic plots of the probe or cavity output power spectral densities ($S_\textrm{c}(\omega)$ or $S_\textrm{a}(\omega)$) vs the deviation of the spectral frequency from the mechanical frequency $\omega-\omega_0$ (s$^{-1}$). 
Case (a) is the torsion pendulum, and the probe field is measured. Parameters: $\omega_0 = 0.2$ s$^{-1}$; $\omega_\textrm{SN} = 0.039$ s$^{-1}$; $M = 10$ kg; $L = 0.1$ m; $Q_\textrm{m} = 5 \times 10^5$; $Q_\textrm{cav} = 10^8$; $P_\textrm{c} = 0.1$ W; $P_\textrm{a} = 5$ mW. The resonance peaks are: $S_\textrm{c}(\omega_0) = 3.6 \times 10^{18}$ and $S_\textrm{c}(\omega_\textrm{sc}) = 5.5 \times 10^{11}$.
Case (b) is LIGO-like, and the cavity output field is measured. Parameters: $\omega_0 = 2 \pi \times 0.6$ s$^{-1}$; $\omega_\textrm{SN} = 0.013$ s$^{-1}$; $M = 40$ kg; $L = 4$ km; $Q_\textrm{m} = 10^6$; $Q_\textrm{cav} = 8 \times10^{12}$; $P_\textrm{c} = 0$ W; $P_\textrm{a} = 700$ W. The resonance peaks are: $S_\textrm{a}(\omega_0) = 3.4 \times 10^{24}$ and $S_\textrm{a}(\omega_\textrm{sc}) = 3.2 \times 10^{22}$. 
Case (c) is the pendulum mode of the levitating mirror, and the probe field is measured. Parameters: $\omega_0 = 10$ s$^{-1}$; $\omega_\textrm{SN} = 0.013$ s$^{-1}$; $M = 0.3$ mg; $L = 0.1$ m; $Q_\textrm{m} = 10^8$; $Q_\textrm{cav} = 10^8$; $P_\textrm{c} = 0.1$ W; $P_\textrm{a} = 10$ mW. The resonance peaks are: $S_\textrm{c}(\omega_0) = 1.9 \times 10^{23}$ and $S_\textrm{c}(\omega_\textrm{sc}) = 7.9 \times 10^{24}$. In all cases: $T_\textrm{s} = 293$ K; $\theta = \pi/2$, the detuning $\Delta = 0$, and the laser wavelength is 1064 nm.
}
\label{combined}
\end{figure}

In this section we apply our simplified optomechanical model to parameters representative of three systems that might be used to test the Schr\"{o}dinger-Newton equation: a torsion pendulum \cite{Harms}, LIGO \cite{Adhikari}, and an optically levitated mirror \cite{Guccione}. For certain parameter ranges the power spectral density defined by Eq.~(\ref{spectral density}) has the distinctive two-peaked structure shown in Fig.~\ref{combined}. The resonance at the lower mechanical oscillator frequency $\omega_0$  is due to the thermal contribution, while that at the semiclassical frequency $\omega_\textrm{sc}$ is due to the quantum contribution \cite{Yang}. Its observation would be evidence for the Schr\"{o}dinger-Newton equation. Since our model is highly simplified, its main purpose is to identify candidate optomechanical systems that might be capable of definitively testing the Schr\"{o}dinger-Newton equation.

The three cases considered in Fig.~\ref{combined} have different distributions of spectral power between the two resonances.  For the torsion pendulum, Fig.~\ref{combined}(a), and the LIGO-like case, Fig.~\ref{combined}(b), the classical resonance dominates.  While for the levitating mirror, Fig.~\ref{combined}(c), the semiclassical resonance dominates.  

If there is no gravitational self-interaction, in which case the Schr\"{o}dinger-Newton theory is wrong, the semiclassical frequency reduces to the mechanical frequency, $\omega_\textrm{sc} = \omega_0$, and the mechanical susceptibility functions are identical, $\chi_\textrm{sc} = \chi_\textrm{m}$. The two resonances then combine into a single resonance having the sum of their spectral power. Hence if the spectral densities in Fig.~\ref{combined} were measured with a frequency resolution insufficient to separate the two resonances, there would be no difference between the results with or without gravitational self-interaction. In Fig.~\ref{combined}(c), for the levitating mirror, the peaks are separated in frequency by $\Delta \omega \approx10^{-5}$ s$^{-1}$, and to separate them would require a measurement time of $2 \pi/ \Delta \omega \approx 6 \times10^5$ s, or about a week. The inverse dependence of the frequency separation on the mechanical frequency, Eq.(\ref{semiclassical frequency approximation}), implies that the required measurement time increases with the mechanical frequency, which is a disadvantage of high frequency systems. Considerations such as these lead us to believe that optomechanical measurements capable of detecting the Schr\"{o}dinger-Newton gravitational self-interaction remain to be made.

A conservative criterion for the visibility of the semiclassical resonance is that at the semiclassical frequency the quantum contribution exceeds the thermal contribution, $S_{i}^\textrm{q} (\omega_\textrm{sc}) > S_{i}^\textrm{cl} (\omega_\textrm{sc})$. For the probe field we give approximate expressions for these contributions for zero cavity detuning, $\Delta = 0$, so that $\Sigma (\omega) = 0$. Provided $k_{c,3} \gg 1$ and $\omega_\textrm{SN}^2 / \omega_\textrm{sc} \gg \gamma_m$ then 
the ratio is,
\begin{equation} \label{SNratio}
\frac{S_c^\textrm{q} (\omega_\textrm{sc})}{S_c^\textrm{cl} (\omega_\textrm{sc})} \approx
\frac{ 2 \omega_l \hbar \omega_\textrm{SN}^4}{\gamma_\textrm{m}^3 \omega_\textrm{sc}^2 c^2 M k_b T}  
\left( P_c +
\left( \frac{2 c}{L \gamma_\textrm{c}} \right)^2 P_a  \right),
\end{equation}
where $P_c$ is the probe laser power and $P_a$ is the input laser power to the cavity. When this is much greater than one the quantum contribution dominates at the semiclassical frequency and the ratio of the peak heights is approximately,
\begin{equation} \label{peakratio}
\frac{S_c^\textrm{q} (\omega_\textrm{sc})}{S_c^\textrm{cl} (\omega_0)} \approx
\frac{\gamma_\textrm{m}^2 \omega_0^2}{\omega_\textrm{SN}^4}
\frac{S_c^\textrm{q} (\omega_\textrm{sc})}{S_c^\textrm{cl} (\omega_\textrm{sc})} .
\end{equation}
These ratios increase with decreasing: temperature, oscillator mass, and mechanical damping. They also increase with laser power. The semiclassical peak visibility, Eq.~(\ref{SNratio}), also increases with decreasing oscillator frequency. 

Fig.~\ref{combined}(a) shows the quadrature power spectral density of the probe output field for parameters consistent with a room temperature torsion pendulum made of steel \cite{Harms}: mass $M = 10$ kg and mechanical frequency $\omega_0 = 0.2$ s$^{-1} $.  According to inequality (\ref{TemperatureInequality}) for the validity of the theory, the center-of-mass motion must be cooled such that $T_\textrm{cm} \ll 0.3$ K. If the bulk mass couples sufficiently weakly to the center-of-mass motion, this does not necessarily mean that the bulk temperature, $T$, must be this cold.

LIGO is an extremely complex optomechanical instrument that makes some of the most precise measurements ever made \cite{Adhikari}. The cavity mirrors are suspended as pendulums with mass $M = 40$ kg and frequency $\omega_0 = 2 \pi \times 0.6$  s$^{-1}$. They are room temperature quartz with estimated Schr\"{o}dinger-Newton frequency $\omega_\textrm{SN} = 0.013$ s$^{-1}$. We optimistically assume a mechanical $Q_\textrm{m} = 10^6$ at this frequency. The optical cavities have length $L = 4$ km and $Q_\textrm{cav} = 8 \times 10^{12}$. Fig.~\ref{combined}(b) shows the quadrature spectral density of the cavity output field, $S_\textrm{a}(\omega)$.  The resonant peaks differ in frequency by $\Delta \omega \approx 2 \times 10^{-5}$ s$^{-1}$, and to separate them would require a measurement time of $2 \pi/ \Delta \omega \approx 3 \times10^5$ s, or about 100 hours.  The inequality (\ref{TemperatureInequality}) only requires that the center-of-mass motion be such that $T_\textrm{cm} \ll 1500$ K. However, the more fundamental criterion is that on the measurement time scale, $\Delta x_\textrm{cm} \ll \Delta x_\textrm{nuc}$. For room temperature silicon, $\Delta x_\textrm{nuc} \approx 6 $ pm. 
Although LIGO data for frequencies near the pendulum resonance are not available, we do not expect that it achieves the required center-of-mass position stability.

Considering the peak visibility, a promising system for testing the Schr\"{o}dinger-Newton equation is a low-mass, low-frequency, low-damping oscillator made of a material with a high Schr\"{o}dinger-Newton frequency. The  levitating mirror is made of quartz and has mass $M=0.3$ mg \cite{Guccione}. It has a high-frequency optical spring mode of about 500 kHz and a low-frequency pendulum mode. The high-frequency mode is in the vertical direction and is due to the stiff optical springs supporting the mirror against the Earth's gravity. In the low-frequency mode the center-of-mass moves like a pendulum along a circular arc about the mirror's center of curvature, while the cavity lengths are unchanged due to the levitating mirror's convex spherical surface \cite{Guccione}. The pendulum mode frequency is determined by the radius of curvature of the mirror, which we assume to be about 10 cm so that  $\omega_0 = 10$  s$^{-1}$. The mechanical  $Q_\textrm{m} = 10^{8}$. Fig.~\ref{combined}(c) shows the quadrature spectral density of the probe output field, $S_\textrm{c}(\omega)$. At the semiclassical frequency it is more than an order of magnitude higher than at the mechanical frequency. The stability criterion is the same as for the LIGO-like case: $\Delta x_\textrm{cm} \ll 6 $ pm over the measurement time of about 30 minutes. The requirement on the center-of-mass temperature of the mode is: $T_\textrm{cm} \ll 40 \mu$K.

\section{Conclusion}
\label{Conclusion}

We have shown that low-frequency optomechanical experiments are able to confirm or refute the many-body Schr\"{o}dinger-Newton equation of Yang \textit{et al.} \cite{Yang}. This is because the gravitational self-interaction separates the resonance at the mechanical frequency into two: one at the mechanical oscillator frequency $\omega_0$, and the other at the higher semiclassical frequency $\omega_\textrm{sc}$. The latter is driven only by the quantum noise of the optical field, whereas the resonance at the mechanical  frequency is driven only by classical noise. The confirming signature would be the observation of these two resonances in the output optical quadrature power spectral variance.  In addition, the semiclassical frequency has a characteristic dependence on the particular material that the oscillator is made of. 

In principle, the Schr\"{o}dinger-Newton equation has no free parameters and so absolute quantitative predictions may be made. However, a significant limitation of our theoretical analysis is the estimation of the position variance of the nuclei $\Delta x^2_\textrm{nuc}$, and hence of the Schr\"{o}dinger-Newton frequencies, using the Debye approximation. The Schr\"{o}dinger-Newton frequencies for compound materials such as quartz are particularly uncertain. 

Among the experimental challenges will be achieving the required high mechanical $Q_\textrm{m}$, and the long observation times needed to separate the mechanical and semiclassical frequencies. The latter imposes severe stability requirements on the experiment, as does ensuring that the center-of-mass modes satisfy the inequality $\Delta x_\textrm{cm} \ll \Delta x_\textrm{nuc}$ for the validity of the approximation to the gravitational self-energy. Nevertheless, our analysis suggests that optomechanical experiments to test the Schr\"{o}dinger-Newton equation may be achievable with existing technology. 

\begin{acknowledgments}
We acknowledge the contributions of Karun Paul, Bram Slagmolen, Yanbei Chen and Rana Adhikari to this work.
\end{acknowledgments}


\begin{thebibliography}{99}
\bibitem{SNfrequencyderivation} This result is given in \cite{Yang} and follows after evaluating the integral for the self-gravitational energy using the Fourier transform method of \cite{Iwe}.
\end{thebibliography}
\end{document}